\documentclass[11pt]{article}

\usepackage[margin=1in]{geometry}
\usepackage{authblk}
\usepackage{amsmath, amssymb}
\usepackage{graphicx}
\usepackage{hyperref}
\usepackage{titlesec}
\usepackage{setspace}
\usepackage{caption}
\usepackage{cite}
\usepackage{xcolor}
\usepackage{adjustbox}
\usepackage{subscript}
\usepackage{tablefootnote}
\usepackage{threeparttable}
\usepackage{booktabs}
\usepackage[labelfont=bf]{caption}
\usepackage[normalem]{ulem}

\titleformat{\section}{\large\bfseries}{\thesection}{1em}{}
\titleformat{\subsection}{\normalsize\bfseries}{\thesubsection}{1em}{}

\title{\textbf{Sample-based Quantum Diagonalization Methods for Modeling the Photochemistry of Diazirine and Diazo Compounds}}

\author[1]{Saurabh Shivpuje\textsuperscript{*}}
\author[2]{Tanvi P. Gujarati\textsuperscript{*}}
\author[1]{Richard Van}
\author[1]{Frank C. Pickard IV}
\author[3]{Triet Friedhoff}
\author[2]{Ieva Liepuoniute}
\author[1]{Wade Davis}
\author[3]{Gavin O. Jones}
\author[1]{Alexey Galda}

\affil[1]{Moderna, Cambridge, MA 02139, USA}
\affil[2]{IBM Quantum, Almaden Research Center, San Jose, CA 95120, USA}
\affil[3]{IBM Quantum, T. J. Watson Research Center, Yorktown Heights, NY 10598, USA}

\date{}

\begin{document}

\maketitle
\begingroup
  \renewcommand\thefootnote{\fnsymbol{footnote}}
  \footnotetext[1]{These authors contributed equally to this work.}
\endgroup
\begin{abstract}
Diazirines and diazo compounds are widely employed as photoreactive precursors for generating carbenes, key intermediates in chemical biology and materials science. However, computationally modeling their reaction pathways remains challenging due to a need for large active spaces and the requirement to accurately capture excited-state surfaces along with transition states and conical intersections. In this work, we utilize a hybrid quantum–classical workflow for investigating carbene formation in representative diazirine–diazomethane systems. Our approach leverages Sample-based Quantum Diagonalization (SQD) and its extended variant (Ext-SQD) for ground and excited-state analysis, combined with classical tools for geometry optimization, active-space selection, and diagnostic evaluation. Quantum computations were carried out on superconducting quantum processors, and results for both aliphatic and aryl-substituted diazirine–diazomethane pairs were benchmarked against established classical methods, including DFT, CCSD, CASCI, and SCI. SQD achieves accuracy surpassing the chemical accuracy threshold for nearly all stationary points on the potential energy surface of parent diazirine relative to the CASCI(12,10) reference, and remains close to chemical accuracy for phenyl-substituted diazirine in a (30,30) active space, with an average deviation of 1.1 kcal/mol relative to the SCI benchmark. SQD closely follows CASCI and SCI trends, showing consistent agreement. The findings demonstrate the promise of quantum computing frameworks in modeling photochemical transformations of electronically complex and pharmacologically relevant molecules. 

\end{abstract}

\section{Introduction}
Nature has long harnessed photons to drive a wide range of chemical transformations, laying the foundation for the field of photochemistry~\cite{nature_photochem1,nature_photochem2}. Through human innovation, this field has evolved into a powerful approach for controlled chemical modification with important applications in medicinal chemistry and it continues to evolve to this day~\cite{photochem_jrny1,photochem_jrny2,photochem_jrny3,photochemistry_jrny4}. One significant development is photoaffinity labeling (PAL), introduced in the 1960s, which has become a key technique in the life sciences by enabling precise covalent tagging of biomolecules to investigate molecular interactions~\cite{first_photolabeling}. Among the photoreactive chemical groups, diazirines are ideal for photolabeling applications due to their ability to cross-link under mild long-wave ultraviolet light~\cite{tpd_longwave}, reducing protein damage, while the ability of the carbene to be rapidly quenched in water~\cite{water_quenching} enhances labeling specificity by favoring covalent attachment only at true binding sites~\cite{diazirine_photoreview,diazirine_drug}.

 Numerous members of the diazirine family have been studied extensively to date; there has been no clear understanding of reaction pathways. The photochemical interconversion between diazirines and diazo compounds, both of which can generate carbenes, proceeds through multiple competing intermediates and pathways, including excited-state rearrangement products~\cite{aliphatic_review, phenyldz_abinitio}, as summarized schematically in Figure~\ref{figure_1}(a). This can be attributed to several factors, including internal isomerization, transition states, the involvement of excited state potential surfaces, and the presence of conical intersection structures~\cite{arenas2002carbene,phenyldz_abinitio,effects_of_substituents}. Experimentally, characterizing transient intermediates is highly challenging~\cite{diazirine_experimental}. On the other hand, modeling these attributes with computational methods is hindered by the complexity of their electronic structure. In particular, the large number of electrons and orbitals (collectively referred to as the active space) required for modeling these systems imposes significant computational demands. Quantum computing algorithms such as Sample-based Quantum Diagonalization (SQD)~\cite{sqd} based on the Quantum Selected Configuration Interaction (QSCI) method~\cite{qsci}, offers a promising avenue for addressing these challenges by enabling accurate estimations of the relevant electronic state energies.

The current accuracy benchmark for performing quantum chemistry calculations using classical computers, referred to as the full configuration interaction (FCI) method, scales factorially for a given number of electrons and orbitals~\cite{fci,fci_review}. To our knowledge, the largest simulation with FCI computed to date is that of propane in the STO-3G basis set describing 26 electrons in 23 orbitals (26e,23o)~\cite{largest_fci}. Recent advances have demonstrated the potential of quantum-centric methods, such as the SQD framework, to provide alternative high accuracy methods by sampling important configurations from a unitary description of the desired eigen-state that can be run using a quantum computer. SQD has been used to study various systems and has shown promising results for systems such as the breaking of the triple bond in the nitrogen molecule (N\textsubscript{2}), predictions of eigenstates for iron-sulfur clusters~\cite{sqd}, solvent-solute interactions~\cite{sqd_solute_solvent}, supramolecular interactions~\cite{sqd_supramolecular}, open-shell system of carbene triplet~\cite{sqd_methylene}, and reaction pathways for polymer degradation~\cite{sqd_polymer}.
Notably, many methods derived from SQD are being developed to widen its range of applications and push its computational limits. These include Extended SQD (Ext-SQD)~\cite{sqd_excited_state}, quantum Krylov states-based approaches~\cite{sqd_kqd,sqd_skqd,sqd_randomized} and the combination of entanglement forging with SQD~\cite{sqd_polymer}. 

Progressing on the established utility of SQD and the inherent demands of diazirine chemistry, this work demonstrates the systematic application of SQD to estimate the energies of chemical species encountered on the potential energy surface (PES) of reactions involving the interconversion of diazirine and diazo compounds, requiring accurate simulations of stationary points (including transition states) on the electronic ground and excited surfaces. We illustrate how the extended formulation of SQD can be effectively employed to access relative excited-state energies. The parent stationary points studied here, represent essential entry points for understanding the photochemistry of more complex and application-relevant members of this molecular family. Furthermore, we highlight the capability of SQD to scale by presenting a 64-qubit SQD calculation of phenyl diazirine, the prototypical aryl diazirine, marking a significant step forward in bringing quantum computational methods to chemically meaningful systems in medicinal chemistry.

The details of this work are organized as follows: Section 2 outlines the methodologies employed, providing the system configurations and a brief review of existing SQD and Ext-SQD approach; Section 3 presents computational results obtained for the two representative systems investigated and discusses their implications. Section 4 concludes the manuscript with a summary of the significant results.  

\section{Methods}
\subsection{Overview of the Computational Workflow}
\begin{figure}[!ht]
    \centering
    \includegraphics[width=0.9\linewidth]{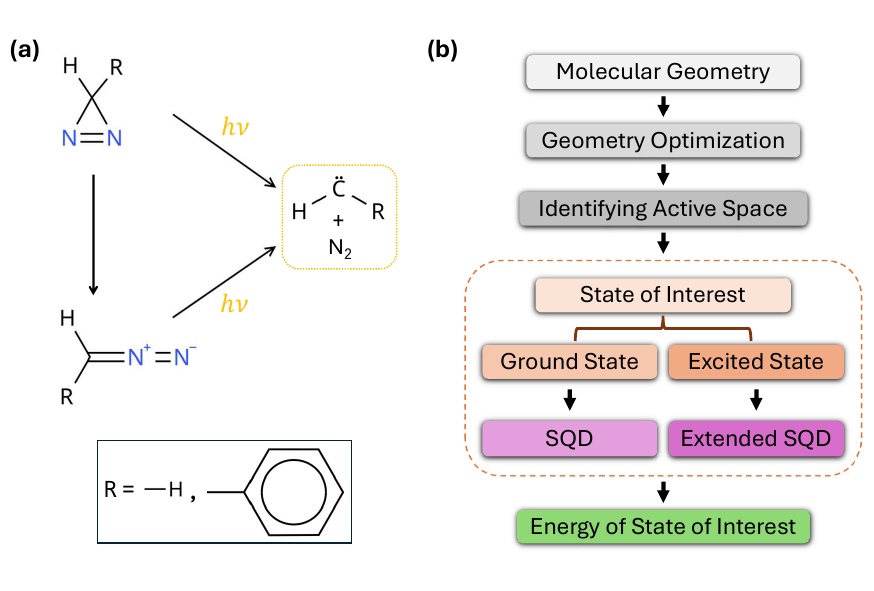}
    \caption{
    (a) Schematic representation of carbene formation upon photolysis of DZ, DM, PhDZ, and PhDM. The diagram provides a simplified overview of the photochemical pathways, highlighting carbene generation from diazirine–diazo pairs and the isomerization of diazirines to diazo compounds.
    (b) Overview of the hybrid quantum–classical workflow employed to compute energies using SQD and extended SQD methodologies.
    }
    \label{figure_1}
\end{figure}

We present a hybrid quantum-classical computing workflow for calculating carbene formation energies in selected candidates from the diazirine and diazoalkane families, see Figure~\ref{figure_1}(a). As illustrated in Figure~\ref{figure_1}(b), the workflow begins with classical computational tools to determine molecular geometries and identify active spaces. We then implement a quantum circuit based on the unitary cluster Jastrow (UCJ) ansatz~\cite{jastrow_wavefunction, jastrow_ansatz, sqd} to sample relevant electron configurations, followed by post-processing to compute energy values. 

Geometry optimization and active space determination were performed using ORCA~\cite{orca_package, orca_review}.  Python-based packages were employed for quantum simulations: molecular integrals and complete active space configuration interaction (CASCI) results were obtained with \texttt{PySCF}~\cite{pyscf2018}, while quantum computations were executed using Qiskit~\cite{qiskit_2019, qiskit-addon-sqd} on the \texttt{ibm\_kingston} quantum processing unit which is one of the Heron r2 processors. The Avogadro~\cite{avogadro_1, avogadro_2} software package was used to plot chemical structures.

\subsection{System Preparation} 
To ensure both chemical diversity and relevance to practical applications, we selected two representative pairs of isomers for a detailed study: (1) the parent compounds 3H-diazirine (DZ) and diazomethane (DM), and (2) their phenyl-substituted analogues, phenyldiazirine (PhDZ) and phenyldiazomethane (PhDM). Selection of these candidates was further guided by the availability of prior experimental and theoretical investigations, which provide a valuable foundation for comparison and validation of our results~\cite{diazirine_experimental,application_photolabeling,effects_of_substituents,phenyldz_abinitio}. In this work, all systems were modeled in the gas phase. Each system was treated as a singlet with a net charge of zero. 

For the parent diazirine system, investigations were conducted based on the reaction profiles described in the work by Arenas et al.~\cite{arenas2002carbene}, discussed in detail in Section \ref{sec:photochem_diaz}. Ground state surface with stationary points for the parent compounds, DM and DZ, along with associated transition states, TS0 and TS1, were considered. The excited state surface included the excited structures corresponding to the DM and DZ systems. In addition, a conical intersection between the ground and excited state surface, CI1, was included in this study. Initial geometries for all the stationary points of DZ–DM pair (except for transition state TS0) were adopted from the prior study~\cite{arenas2002carbene}. We did not use the reported TS0~\cite{arenas2002carbene} because that structure was not stable under our chosen active space, and their reported vibrational frequency analysis did not include the requisite imaginary frequency for a transition state; accordingly, we derived and verified our own TS0, and we have provided the corresponding coordinates and vibrational frequency data in the Supplementary Information. 

The active space was selected based on the electronic factors governing its photochemistry. In particular, the molecular orbitals contributing to the lowest-lying excited state, which is central to carbene generation, were considered. Time-Dependent Density Functional Theory (TD-DFT)~\cite{tddft} calculations indicated that the HOMO and LUMO orbitals capture the essential orbital contributions to this excitation. In the case of the parent DZ and DM, orbitals involved in the first electronic excitation were computed with TD-DFT/cc-pVTZ using the B3LYP functional. Nevertheless, in order to ensure both a reliable description of electron correlation and a sufficiently large active space to meaningfully assess the performance of SQD relative to conventional approaches, we extended the orbital set beyond the minimal configuration. Specifically, a 12-electron, 10-orbital (12,10) active space was adopted. Geometry optimizations at the CASSCF/cc-pVTZ level with this active space were carried out using ORCA~\cite{casscf} to further refine the structures and enable determination of the TS0 state. To identify the transition state (TS0) associated with the isomerization between diazirine (DZ) and diazomethane (DM) on the ground-state surface, a scan was performed by interpolating structures along the reaction coordinate defined by the angle between the C--N and N--N bonds. Transition-state optimizations were then performed with the OPTTS keyword enabled in ORCA. Excited-state geometry optimizations for the parent DZ and DM systems were performed at the state-averaged CASSCF(12,10)/cc-pVTZ level. The conical intersections were located using the CI-OPT functionality available at the TD-DFT/cc-pVTZ level (rather than CASSCF). Stationary points, including transition states, were verified through vibrational frequency analysis to ensure stability and the correct number of imaginary modes.

For the phenyl-substituted diazirine (PhDZ) and diazomethane (PhDM) systems, using the geometries reported by Yunlong et al.~\cite{phenyldz_abinitio} as initial structures, ground-state geometries were optimized using the RI-MP2 method~\cite{rimp2,mp2} with the RIJCOSX approximation, in conjunction with the cc-pVTZ basis set. The auxiliary basis sets def2/J and cc-pVTZ/C were employed for the Coulomb and MP2 correlation integrals, respectively. The strategy for defining the active space followed the same principles as for the parent diazirine. We used the MP2 Natural Orbitals to construct the active space instead of the Hartree-Fock based Molecular Orbitals as in the case of the parent system. To demonstrate the scalability of the workflow, we employed a 30-electron, 30-orbital (30,30) active space. This choice ensured the inclusion of all orbitals relevant to the low-lying excitations while simultaneously accounting for extended conjugation effects from the phenyl substituent. 

It is important to note that this study focuses exclusively on the carbene formation pathway originating from the DZ form for parent diazirine and both PhDZ and PhDM for phenyl-substituted diazirine, in keeping with previous studies~\cite{arenas2002carbene, phenyldz_abinitio}. While alternative pathways involving carbene formation from the stable DM form have been proposed in other studies, they are beyond the scope of the present work~\cite{aliphatic_review,phenyldz_abinitio,diazirine_experimental}.

\subsection{Classical  Benchmarking}

The classical computational workflow was extended beyond initial geometry optimization and active-space selection to include energy evaluations using a variety of electronic structure methods. All energy evaluations were performed on optimized geometries, employing the correlation-consistent triple-zeta (cc-pVTZ) basis set 
throughout. Energy calculations were carried out with widely used density functional theory (DFT/B3LYP)~\cite{dft} and correlated methods including coupled cluster with singles and doubles (CCSD)~\cite{ccsd} with the frozen core approximation. For determination of excited state energies, TD-DFT and Equation-of-Motion (EOM) CCSD were used. For the primary comparison of methods, CASCI calculations using the (12,10) active space for the DZ and DM molecules were performed, supplemented with DFT, CCSD and CASSCF comparisons. Excited state energies were computed in an active space obtained by performing state-averaged (SA) CASCI(12,10) and CASSCF calculation with equal weights given to the ground and the first excited state for the structures optimized on the excited state surface. Ground state energies of the phenyl-substituted derivatives were obtained using the selected configuration interaction (SCI)~\cite{sci} approach implemented in \texttt{PySCF} with a significantly larger (30,30) active space. 

\subsection{Sample-based Quantum Diagonalization (SQD) and Ext-SQD}

SQD~\cite{sqd, qsci} is a hybrid quantum–classical algorithm designed to approximate low-energy eigenstates of the electronic structure Hamiltonian. The method works by projecting the Hamiltonian onto reduced subspaces of the full Hilbert space defined by computational basis states sampled from unitary ansatzes implemented on a quantum computer.

In this work, a parameterized quantum circuit representing the desired eigenstate, $\Psi(\vec{\theta})$, is constructed based on the local approximation of the UCJ ansatz~\cite{jastrow_ansatz}. Based on the protocol from Ref.~\cite{sqd}, a circuit with $M$ qubits (which corresponds to the number of spin-orbitals in the system studied) is measured $N_s$ times (a.k.a. shots) in the computational basis. The measured outcomes produce a set of bitstrings, $X = \{x_i\}$ with $x_i \in \{0,1\}^M$, which represent Slater determinants drawn from the probability distribution $P_\Psi(x) = |\langle x | \Psi(\vec{\theta}) \rangle|^2$.

On pre-fault-tolerant quantum devices, noise distorts the ideal distribution $P_\Psi(x)$ into a broadened distribution $\tilde{P}_\Psi(x)$, producing a noisy configuration set $\tilde{X}$. To enforce physical symmetries such as fixed particle number and spin, a self-consistent configuration recovery (S-CORE) routine is applied to filter the raw samples and yield a corrected set $X_R \subseteq \tilde{X}$. From $X_R$, we draw $K$ batches of configurations $\{S^{(k)}\}_{k=1}^K$, each batch containing up to $d$ unique bitstrings, where $d$ is referred to as the samples per batch. Each bitstring encodes a specific electronic configuration: the first half of the bits corresponds to alpha-spin orbitals, and the second half to beta-spin orbitals.  

Each batch $S^{(k)}$ defines a $D$-dimensional subspace of the Fock space, constructed by taking all possible combinations of the unique alpha- and beta-spin bitstrings in that batch. The Hamiltonian is classically projected onto this subspace and diagonalized using the Davidson algorithm, yielding approximate eigenpairs $(|\psi^{(k)}\rangle , \, E^{(k)})$. Spin-orbital occupations $(n_{p\sigma} = \langle \psi^{(k)} | \hat{n}_{p\sigma} | \psi^{(k)} \rangle)$ are averaged over batches and used in the next S-CORE update. The process of recovery, resampling, and diagonalization is repeated for a fixed number of $I_{\text{S-CORE}}$ SQD iterations or until convergence. For further details on the SQD approach, algorithm, and implementation aspects, we refer the reader to Ref.~\cite{sqd}.

Ext-SQD~\cite{sqd_excited_state} generalizes SQD by expanding the configuration space with additional Slater determinants generated via single and/or double excitations from a subset of bitstrings in $X_R$. This subset is determined by performing a SQD calculation to get an approximate ground state and then filtering the bitstrings with magnitudes greater than a specified threshold from this state. This results in an extended basis, denoted by $X_E$, which is then used for diagonalization. The enriched configuration space obtained via this extension improves the quality of the obtained ground and excited state approximations. Further details and algorithmic justification for the Ext-SQD method are given in Ref.~\cite{sqd_excited_state}.

\subsection{Quantum Circuit Construction}

The effectiveness of SQD and its extensions relies on accurate quantum circuit construction to sample the most relevant electron configurations. The quantum-centric pipeline begins with computing the full molecular Hamiltonian:
\begin{equation}
\hat{H} = \sum_{pr\sigma} h_{pr} \, \hat{a}^\dagger_{p\sigma} \hat{a}_{r\sigma} 
+ \sum_{pqrs\sigma\tau} \frac{(pr|qs)}{2} \, 
\hat{a}^\dagger_{p\sigma} \hat{a}^\dagger_{q\tau} \hat{a}_{s\tau} \hat{a}_{r\sigma} \, ,
\label{eq:hamiltonian}
\end{equation}
where the integrals $h_{rs}$ and $(pq|rs)$ are computed using \texttt{PySCF}.

To encode this fermionic Hamiltonian on quantum hardware, we use the Jordan–Wigner transformation~\cite{jw_transformation}, which maps each spin-orbital to a qubit. Consequently, for a chosen active space, the total number of required qubits equals 
the number of spin-orbitals. On QPU's topology, we allocate two parallel chains of qubits for spin-up ($\alpha$) and spin-down ($\beta$) spins, separated by ancilla qubits to allow for $\alpha\beta$ interactions.

Each circuit is initialized in the Hartree–Fock reference state using the built-in functionality contained in Qiskit followed by the Local Unitary Cluster Jastrow (LUCJ) ansatz, generated using the \texttt{ffsim}~\cite{ffsim} package. The parameters for the LUCJ ansatz are derived from the CCSD $T_2$ amplitudes calculated for the same active space~\cite{sqd}. This parameterization provides a good guess for the ansatz parameters and allows sampling of important electronic configurations. The transpilation of circuits for execution on the quantum hardware is carried out using the \texttt{PassManager} function in Qiskit, ensuring device-optimized implementation~\cite{qiskit2024}. To mitigate decoherence during quantum circuit execution, dynamical decoupling was enabled as part of the error mitigation strategy~\cite{viola1998dynamical, kofman2001universal}. 

\section{Results and Discussion}

\subsection{Diazirine and Diazomethane}
\label{sec:photochem_diaz}
DZ-DM were selected as prototypical systems to validate our workflow due to their small size, well-characterized photochemistry, and structural relevance to a broader class of photoactive compounds. Both compounds contain a diazo moiety, which undergoes isomerization and potential nitrogen extrusion to form carbene intermediates of interest (Figure~\ref{figure_1}(a)). These features make them ideal candidates for studying light-induced reaction pathways and benchmarking quantum simulation methodologies.

\begin{figure}[!ht]
  \centering
  \fbox{%
    \includegraphics[width=0.9\linewidth]{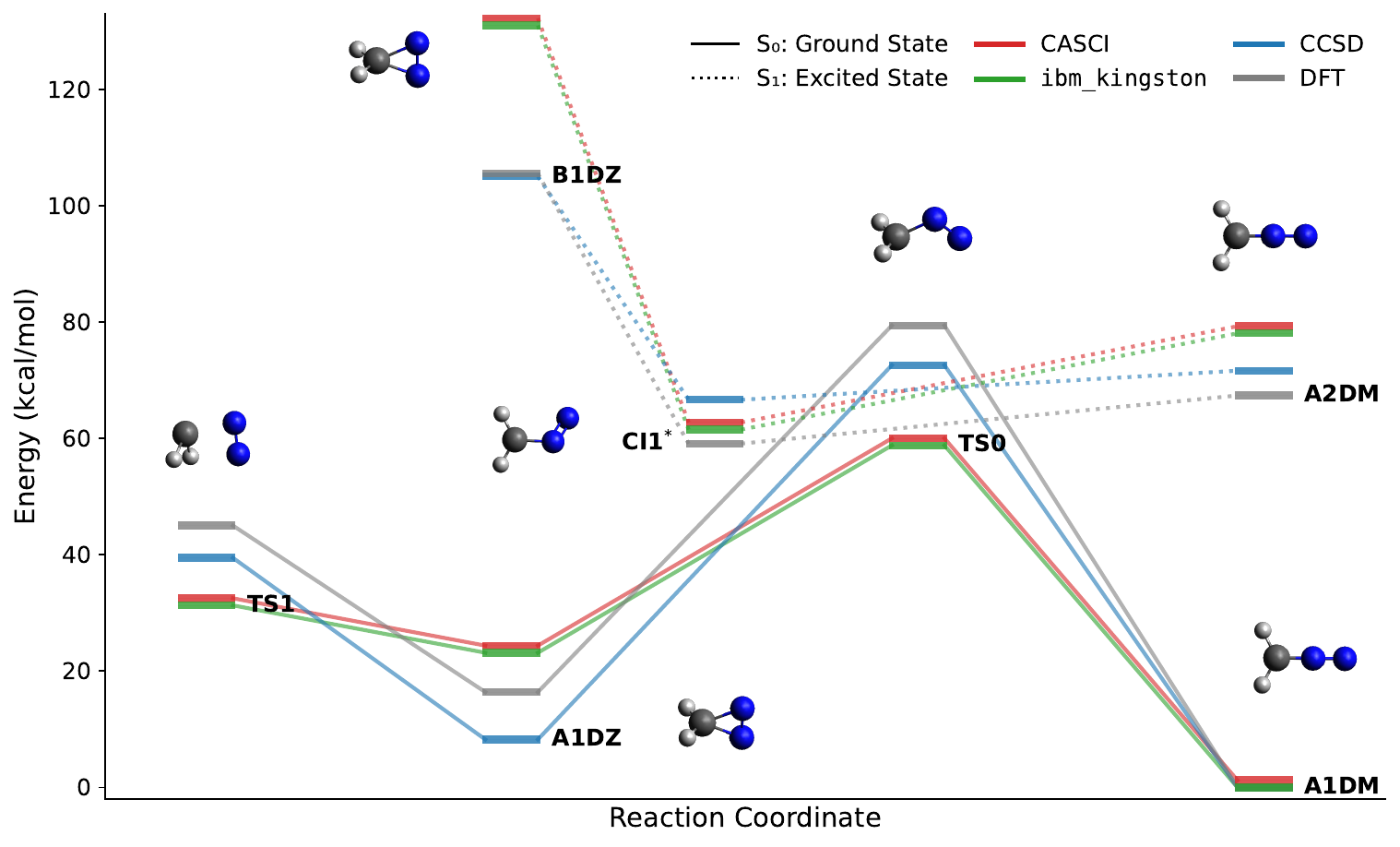}%
  }
  \caption{Relative energies (in kcal/mol) of stationary points along the reaction coordinate for the DZ – DM interconversion. PES were computed with DFT, CCSD, CASCI(12,10) and SQD(12,10) on \texttt{ibm\_kingston}. Points span both S$_0$: ground and S$_1$: excited states. Geometries at representative points along the reaction path are illustrated in the plot for reference. In the molecular model, atoms are color-coded as follows: carbon in gray, hydrogen in white, and nitrogen in blue. The \texttt{ibm\_kingston} markers are shown slightly shifted vertically for clarity; in fact, they overlap very closely with CASCI values. $^{*}$Note that the value of the conical intersection, CI1, was computed using the CCSD2 implementation in ORCA.
}
  \label{fig:diazirine-pes}
\end{figure}

The photochemistry of parent DZ-DM is largely governed by processes initiated between the ground state and the first excited states~\cite{arenas2002carbene}, accordingly we focused our analysis on seven key stationary points on the ground ($S_0$) and first excited ($S_1$) PES, as illustrated in Figure~\ref{fig:diazirine-pes}. These include two ground-state minima: 3H-Diazirine (A1DZ) and Diazomethane (A1DM); two transition states: TS0, corresponding to the isomerization barrier between A1DZ and A1DM, and TS1, associated with the dissociation of A1DZ into carbene and nitrogen; one conical intersection (CI1) connecting the ground and first excited states; and two excited-state minima: the first excited state of DZ (B1DZ) and the first excited state of DM (A2DM). When diazirine (A1DZ) or diazomethane (A1DM) absorb light, they are excited to their higher-energy states (B1DZ and A2DM, respectively).

Key structural parameters, including bond lengths and angles determined in this study, demonstrate excellent agreement with previously reported experimental values~\cite{dm_experimental_structure, dz_experimental_structure}. These parameters are summarized in Table~S1 of the Supplementary Information. Optimized atomic coordinates and vibrational frequencies for each of these structures are also provided in the Supplementary Information.

Electronic energies were determined by using the electronic configurations sampled from the LUCJ ansatz with a single layer on the \texttt{ibm\_kingston} quantum processor. The (12,10) active space was encoded onto a 20-qubit register, arranged as two parallel chains of 10 qubits each with 3 ancilla qubits for inter-chain connectivity. SQD calculations were carried out using $N_s = 100{,}000$ shots, $I_{\text{S\\-CORE}} = 5$ iterations and $K = 50$ batches, each containing $d = 500$ unique samples. Figure~\ref{fig:diazirine-pes} shows benchmark results for ground (excited) state obtained with DFT (TD-DFT), CCSD (EOM-CCSD), CASSCF(SA-CASSCF) and CASCI(SA-CASCI) in the active space, compared against SQD(Ext-SQD) relative energies; for the excited state, the reported energy corresponds to the first excited root. For the case of the conical intersection CI1, the CCSD energy was computed with the CCSD2 method in ORCA, a lower-scaling variant available for closed-shell systems. This was done because the default CCSD calculation did not converge for this system. For Ext-SQD method all single and double excitations of SQD eigen-state configurations with co-efficient magnitudes greater than 0.01 were included in the set $X_E$.

We first describe the PES using CASCI(12,10) as a consistent multireference benchmark, serving as a fully classical counterpart to SQD while maintaining the same active space. On the S$_0$ surface, A1DM is identified as the global minimum, with A1DZ lying higher by $\sim$23 kcal/mol (Table~\ref{tab:energy_barriers_DM_DZ}). The barrier TS0 appears at $\sim$59 kcal/mol relative to A1DM, while the carbene formation transition state TS1 occurs at $\sim$31 kcal/mol. On S$_1$, the minima A2DM and B1DZ are located at $\sim$78 and $\sim$131 kcal/mol, respectively. The S$_1$/S$_0$ conical intersection CI1 lies at $\sim$59 kcal/mol, slightly above TS1. In line with established photochemistry, vertical excitation from A1DM or A1DZ populates the corresponding S$_1$ wells, followed by relaxation through CI1, which directs reactivity toward carbene formation through TS1.

DFT and CCSD calculations reveal notable differences. On the S$_0$ surface, the two methods differ by $\sim$6 kcal/mol on average, with CCSD predicting lower relative energies. On S$_1$, the discrepancy decreases to $\sim$3.6 kcal/mol, though CCSD predicts higher energies for A2DM and CI1 than DFT. These inconsistencies emphasize the inherent challenge of describing this PES with single-reference methods. Comparison between CASSCF and CASCI further shows that the DZ and S$_1$ states are especially sensitive to orbital relaxation, highlighting how methodological choices significantly affect predicted energetics. Relative to the CASCI(12,10) benchmark (Figure~\ref{fig:diazirine-pes}), SQD provides excellent agreement within the same (12,10) active space. Energies for all stationary points deviate by less than 1 kcal/mol, except for CI1, which differs by only 3 kcal/mol. Such accuracy at chemically distinct stationary points on the photochemical PES demonstrates the robustness of the SQD approach.

\setlength{\fboxsep}{0pt}
\setlength{\fboxrule}{0.5pt}

\begin{table}[ht!]
\centering
\caption{\textbf{Computed energy barriers (kcal/mol) for DZ-DM structures using various quantum chemistry methods, including comparisons with previous studies.}}
\begin{threeparttable}
\resizebox{\textwidth}{!}{
\begin{tabular}{lrrrrrr}
\toprule
\multicolumn{7}{c}{\textbf{DZ-DM}} \\
\midrule
Structure  & DFT & CCSD & CASSCF \small(12,10)\tnote{a} & CASSCF \small(12,10) & CASCI \small(12,10) & SQD \small(12,10) \\
\midrule
A1DM  & 0.00 & 0.00 & 0.00 & 0.00 & 0.00 & 0.00 \\
A1DZ   & 16.39 & 8.24 & 12.30 & 8.32 & 23.16 & 23.16 \\
TS0    & 79.44 & 72.62& 56.97 & 60.91 & 58.81 & 58.82 \\
TS1   & 45.02 & 39.47& 36.10 & 30.57 & 31.30 & 31.30 \\
\midrule
Structure & TD-DFT & EOM-CCSD & CASSCF \small(12,10)\tnote{a} & SA-CASSCF \small(12,10) & SA-CASCI \small(12,10) & Ext-SQD \small(12,10)\\
\midrule
A2DM  & 67.40 & 71.66  & 73.49 & 71.07 & 78.10 & 78.09 \\
B1DZ   & 105.55 & 105.07  & 115.44 & 114.40 & 131.05 & 131.04 \\
CI1   & 59.12 & 66.67\tnote{b}  & 77.58 & 50.74 & 59.04 & 61.53 \\
\bottomrule
\end{tabular}
}
\begin{tablenotes}
\tiny
\item [a] CASSCF(12,10) data from Arenas et al.~\cite{arenas2002carbene}. Note that the active space differs from the one used in the present study.
\item [b] CCSD value for CI1 obtained using the CCSD2 implementation in ORCA.
\end{tablenotes}
\end{threeparttable}
\label{tab:energy_barriers_DM_DZ}
\end{table}

Based on the calculated energy differences in this work, A1DZ is expected to convert preferentially either to the diazo compound via the excited state (B1DZ) followed by relaxation to A1DM, or to dissociate directly into carbenes through TS1. The reverse transformation from diazo to diazirine is highly unfavorable due to the substantial energy difference, making a true isomerization equilibrium unlikely. Upon photoexcitation diazo compounds similar to the behavior of diazirine can access higher-energy states and because the energy barrier to TS1 is relatively low, the excess vibrational energy gained during relaxation through the conical intersection should channel the system toward carbene formation rather than reversion to the stable diazirine. This result is consistent with mechanisms previously proposed by theoretical studies~\cite{parentDZ_1994,arenas2002carbene} and confirmed by recent experimental investigation~\cite{diazirine_experimental}.

One of the promising features of SQD is that it is possible to get an accurate description of the desired eigenstate using a fraction of all possible electronic configurations in an active space. For the computations performed on the ground and excited state surfaces, we see that SQD is able to get accurate energies (within micro Hartree of the CASCI value) with a subspace of about 29,000–34,000 determinants, compared with approximately 44,000 determinants required for the CASCI treatment. Absolute energies and SQD parameters, including two-qubit gate counts, are provided in the Supplementary Information.

The reliability of SQD as a multi-reference method is further evident in challenging regions of the PES. At the transition state TS0, where strong multireference character is present ($T_1$ diagnostic value = 0.067 $>$ 0.020~\cite{t1-diagnostic}), conventional single-reference methods show significant deviations from the CASCI benchmark and CASSCF value. SQD maintains close agreement with multireference CASCI method, demonstrating its effectiveness beyond equilibrium structures.

\subsection{Phenyldiazirine and Phenyldiazomethane}

Although aliphatic diazirines are often preferred for their compact size and minimal steric disruption in biological contexts, several studies have demonstrated that aromatic diazirines generate higher yields of reactive carbenes upon photoactivation, making them more effective for covalent capture in labeling applications~\cite{phenyldz_abinitio}. Phenyldiazirine (PhDZ) and Phenyldiazomethane (PhDM) were thus selected as representative aryl systems to test the scalability and versatility of our quantum-classical workflow. 

Four stationary points along the ground-state PES were identified and fully characterized: the minima corresponding to the reactant structures (PhDM and PhDZ), and two distinct transition states (PhTS0 and PhTS1), which define alternative decomposition pathways leading to reactive carbene formation~\cite{phenyldz_abinitio}. Each pathway originates from its respective precursor structure, PhDM or PhDZ. The optimized geometries (Figure S1, provided in Supplementary Information) show excellent agreement with previously reported values~\cite{phenyldz_abinitio}. Complete sets of optimized Cartesian coordinates and vibrational frequencies for all four stationary points are provided in the Supplementary Information.

SQD calculations were performed with a single layer of the LUCJ ansatz sampled on the \texttt{ibm\_kingston} device. The selected (30,30) active space constructed with the MP2 Natural Orbitals was mapped onto 60 qubits with 3 additional ancilla qubits. The number of two-qubit gates in the employed quantum circuit is provided in the Supporting Information. For the SQD calculations, $N_s = 100{,}000$ measurement shots and $I_{\text{S\\-CORE}} = 5$ iterations were used. First, SQD calculations were run with $d = 5*10^{4}$ samples per batch but restricting the maximum subspace dimension for diagonalization to $D = 2.25*10^{8}$. To further improve the results obtained from the SQD computation, Ext-SQD was implemented with additional configurations generated via single excitations on SQD bitstrings with coefficient with magnitudes greater than 0.001. A further diagonalization was performed with the extended set of bitstrings restricting the maximum subspace dimension to $D = 2.25*10^{8}$ due to compute limitations. The lowest energies obtained from SQD and Ext-SQD for individual systems are reported. 

\begin{figure}[!ht]
  \centering
  \fbox{%
    \includegraphics[width=0.9\linewidth]{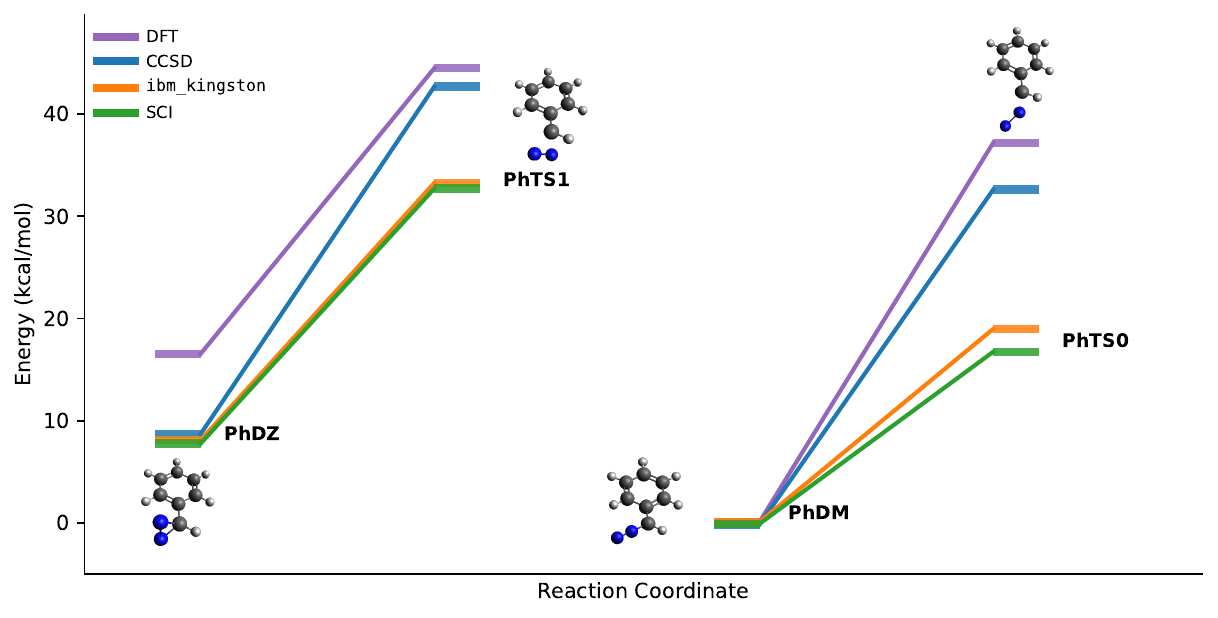}%
  }
\caption{Computed ground-state energy profile for phenyldiazirine (PhDZ) - phenyldiazomethane (PhDM) system along the reaction coordinate, comparing results from quantum simulations and classical methods. Geometries at representative points along the reaction path are illustrated in the plot for reference. In the molecular model, atoms are color-coded as follows: carbon in gray, hydrogen in white, and nitrogen in blue.}
\label{fig:PhDZ-pes}
\end{figure}

The (30,30) active space was selected to capture the $\pi$-system of the phenyl ring along with the frontier orbitals of the diazirine core, as these orbitals dominate the key excitations governing the photoreactivity of aryl diazirines. For benchmarking, classical SCI and CCSD calculations were performed within the same active space using \texttt{PySCF}. In addition, DFT and CCSD energies were also computed. The computed ground-state potential energy profiles for PhDZ and PhDM are presented in Figure~\ref{fig:PhDZ-pes}. Numerical comparison of various methods listed above along with relative energies reported in a prior study by Yunlong et al.~\cite{phenyldz_abinitio} are provided in Table~\ref{tab:energy_barriers_PhDM_PhDZ}. For an active space of size (30,30) since it is not possible to get CASCI energies, we present comparison of the SQD approach with two post Hartree Fock classical methods, CCSD and SCI. We see that SQD energies are within $\sim$1.1 kcal/mol of the SCI relative energies on average. The CCSD relative energies on the other hand differ by $\sim$4 kcal/mol on average with respect to the SQD values. Compared to the active space methods, the full space methods show larger variability in the obtained relative energies. Single reference method DFT differ by $\sim$12 kcal/mol on average, with significant discrepancy between the relative energies for the PhTS0 transition state. We see a larger difference of about 8-12 kcal/mol on average between the relative energies for the full space (CCSD) and the active space methods (CCSD, SCI, SQD), with PhTS0 again showing largest differences. The above trends showcase the complexities of this PES with the need for accurate multi-reference methods to describe the associated relative energies. Optimized atomic coordinates, vibrational frequencies, absolute energy values, and SQD parameters (including two-qubit gate counts) are provided in the Supplementary Information.

\setlength{\fboxsep}{0pt}
\setlength{\fboxrule}{0.5pt}

\begin{table}[ht!]
\centering
\caption{\textbf{Computed energy barriers (kcal/mol) for PhDZ-PhDM structures using various quantum chemistry methods, including comparisons with previous studies.}}
\begin{threeparttable}
\resizebox{\textwidth}{!}{
\begin{tabular}{lrrrrrrr}
\toprule
\multicolumn{7}{c}{\textbf{PhDZ-PhDM}} \\
\midrule
Structure  & DFT & CCSD & RI-CC2/TZVP\tnote{c} & CCSD (30,30) & SCI (30,30) & SQD (30,30) \\
\midrule
PhDM  & 0.00 & 0.00 & 0.00 & 0.00 & 0.00 & 0.00 \\
PhDZ  & 16.67 & 8.55 & 13.30\tnote{b} & 3.94 & 7.65 & 8.20 \\
PhTS0   & 37.31 & 32.61 & 33.94 & 14.85 & 16.83 & 19.01 \\
PhTS1    & 44.69 & 42.82 & 41.96 & 29.97 & 32.56 & 33.06 \\
\bottomrule
\end{tabular}
}
\begin{tablenotes}
\tiny\item [a] RI-CC2/TZVP data from Yunlong et al.~\cite{phenyldz_abinitio}
\item [b] Value as reported in main text of Yunlong et al.~\cite{phenyldz_abinitio}; their supplementary information lists different value.
\end{tablenotes}
\label{tab:tablefootnote}
\end{threeparttable}
\label{tab:energy_barriers_PhDM_PhDZ}
\end{table}

The relative energies obtained with the SQD method appear competitive for phenyl-substituted systems in a active space of size (30,30). This  was achieved with a sub-space of $2.25 \times 10^{8}$ determinants when compared to the size of the full configuration space, $2.41 \times 10^{16}$, within the active space.
The SQD results can be further improved by performing a larger diagonalization with more compute power. In addition, the quality of the sampled determinants can be improved by utilizing an ansatz that captures more electronic correlations. For example, within the LUCJ family, number of LUCJ layers can be increased or circuit parameters that are optimized to lower the electronic energy can be used. The constraints on the locality of the ansatz can be lifted by considering alternative hardware qubit connectivity. A particular mention of the effect of hardware noise is relevant in the context of improving the quality of the sampled bitstrings. For the parent system it was observed that noise present in the quantum hardware helped in improving the diversity of sampled bitstrings. This helped in negating the effect of sampling from a highly peaked distribution around the Hartree-Fock bitstring when using a single layered LUCJ ansatz. In contrast, the phenyl-substituted system showed the opposite trend. Because the circuits used were much larger than those of the parent system, accumulated device noise had a negative impact on the sample distribution. To improve results in these larger active spaces, more advanced error-mitigation protocols can be employed.

\section{Conclusion}

This study establishes SQD as a powerful hybrid quantum–classical framework for exploring complex potential energy surfaces in photochemically relevant systems. Using diazirine and its phenyl-substituted derivatives as representative cases, we show that SQD achieves accuracy comparable to high-level multireference methods while scaling more favorably with increasing active space size, thereby reducing computational demands.

A central outcome of this study is the ability of SQD to probe the multifaceted diazirine–diazo potential energy landscape, a system of particular importance in medicinal chemistry due to its application in photoaffinity labeling. Within this framework, the standard SQD formulation proves particularly effective for describing ground-state energetics, whereas the Ext-SQD variant enables accurate treatment of excited states characterized by strong configuration mixing. This complementary capability highlights the adaptability of quantum–classical strategies across different electronic regimes. We have further demonstrated the scalability of the method by treating phenyl-substituted derivatives with active spaces as large as (30,30), underscoring the efficiency of SQD in handling electronically demanding systems.

SQD emerges as a scalable and reliable tool for advancing electronic structure theory in photochemistry, with direct benefits for medicinal chemistry. This approach enables the calculation of accurate ground- and excited-state surfaces at reduced computational cost, making it a promising tool for photochemical studies and drug discovery. Looking ahead, as quantum hardware advances, there is clear scope for even greater accuracy and for designing improved ansatz that enhance expressibility for larger molecular systems.

\section*{Acknowledgments}
The authors would like to thank David Cascio for his valuable technical assistance in managing the computational resources of this work and Mariana LaDue for her assistance and support with project logistics.

\bibliographystyle{unsrt}
\bibliography{references}

\end{document}


\begin{center}
    {\LARGE \textbf{Supplementary Information}} \\[1em]
\end{center}

\section{Optimized Structures and Vibrational Frequencies}
\subsection{Structure Parameters}
\begin{table}[ht]
\centering
\caption{\textbf{Key bond distances and bond angles for the structures of DZ and DM, reported alongside experimental values. Bond distances are in \AA\ and angles in degrees ($^\circ$).}}
\begin{tabular}{lccccc}
\hline
 & C$_1$--N$_2$ & N$_2$--N$_3$ & C$_1$--H$_4$ & C$_5$--N$_2$--N$_3$ & H$_4$--C$_1$--H$_5$ \\
\hline
Experimental\textsuperscript{\cite{dz_experimental_structure}} DZ & 1.482 & 1.228 & 1.090 & 65.5 & 117.0 \\
A1DZ, ground state DZ & 1.500 & 1.205 & 1.068 & 66.3 & 120.8 \\
B1DZ, 1st excited state DZ & 1.544 & 1.220 & 1.067 & 66.7 & 121.9 \\
Experimental\textsuperscript{\cite{dm_experimental_structure}} DM & 1.320 & 1.120 & 1.080 & 180.0 & 127.0 \\
A1DM, ground state DM & 1.293 & 1.127 & 1.086 & 180.0 & 125.0 \\
A2DM, 1st excited state DM & 1.304 & 1.156 & 1.097 & 180.0 & 121.0 \\
\hline
\end{tabular}

\label{tab:dz_geom_parameters}
\end{table}

\begin{figure}[!ht]
    \centering
    \caption{\textbf{Optimized geometries of (a) phenyldiazirine (PhDZ) and (b) phenyldiazomethane (PhDM) computed at the RI-MP2/cc-pVTZ level. Bond lengths are shown in Å.}}
    \includegraphics[width=0.8\linewidth]
    {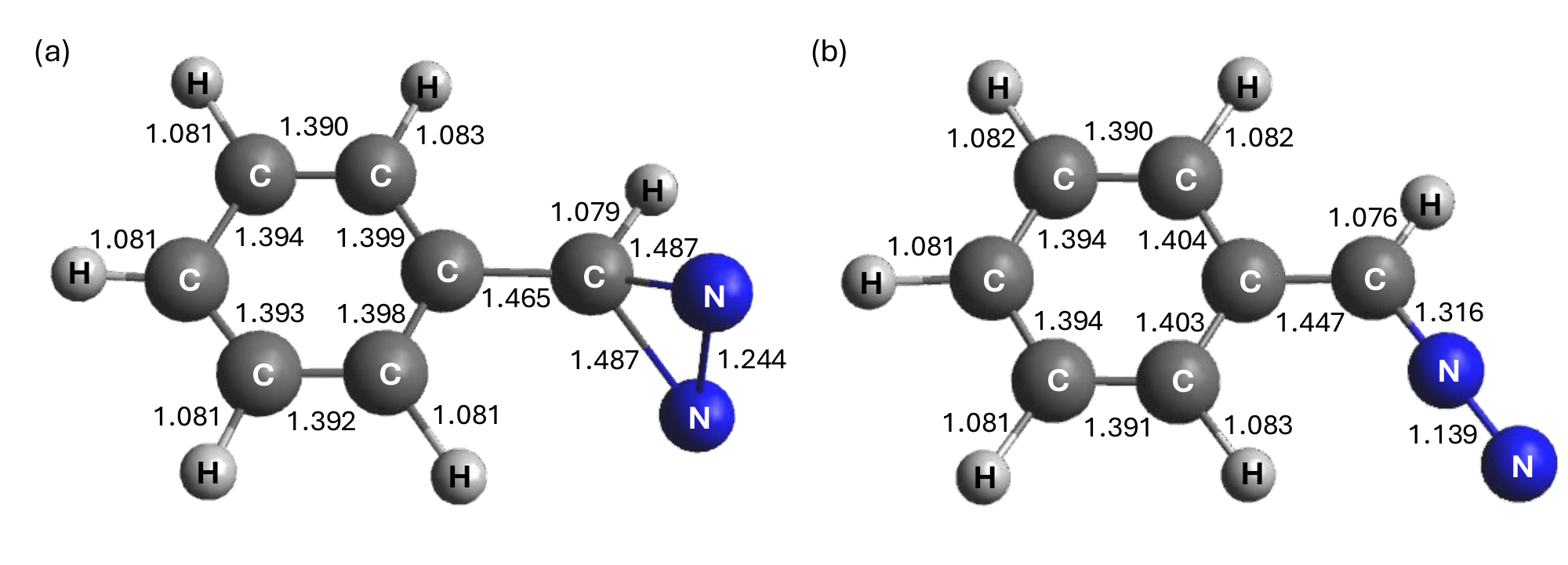}

    \label{fig:phdz_phdm_struct}
\end{figure}
\newpage
\subsection{Structure: A1DM}
\textbf{Level of Theory:} CASSCF (12,10)


\begin{center}
    \includegraphics[trim={5cm 0 0 0},clip,width=0.4\textwidth]{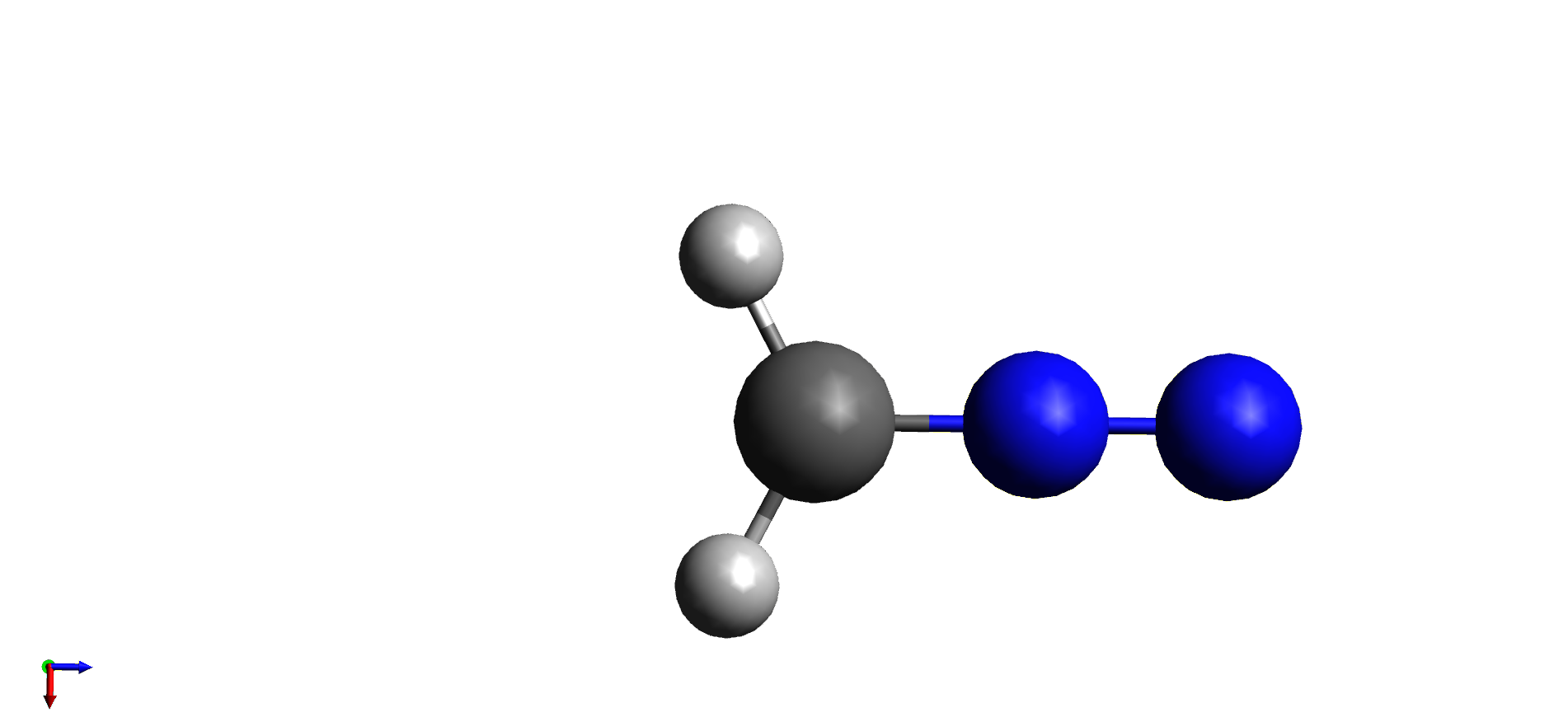}
\end{center}

\subsubsection*{Optimized Atomic Coordinates in Å}
\begin{tabular}{lrrr}
Element & x & y & z \\

C & 0.00000000000000 & 0.00000000000005 & -1.21552957727341 \\

N & 0.00000000000001 & -0.00000000000000 & 0.07724336675476 \\

N & -0.00000000000000 & -0.00000000000001 & 1.20378129220906 \\

H & 0.96390796714505 & -0.00000000000002 & -1.71682954084520 \\

H & -0.96390796714506 & -0.00000000000002 & -1.71682954084521 \\

\end{tabular}

\subsubsection*{Vibrational Frequencies (cm$^{-1}$)}
\begin{tabular}{ cccccccc }

    0.00 & 0.00 & 0.00 & 0.00 & 0.00 & 0.00 & 296.62 & 441.73 \\

    570.09 & 1176.04 & 1229.72 & 1467.46 & 2266.35 & 3103.62 & 3232.80 \\

\end{tabular}

\newpage
\subsection{Structure: A1DZ}
\textbf{Level of Theory:} CASSCF (12,10)


\begin{center}
    \includegraphics[trim={5cm 0 0 0},clip,width=0.4\textwidth]{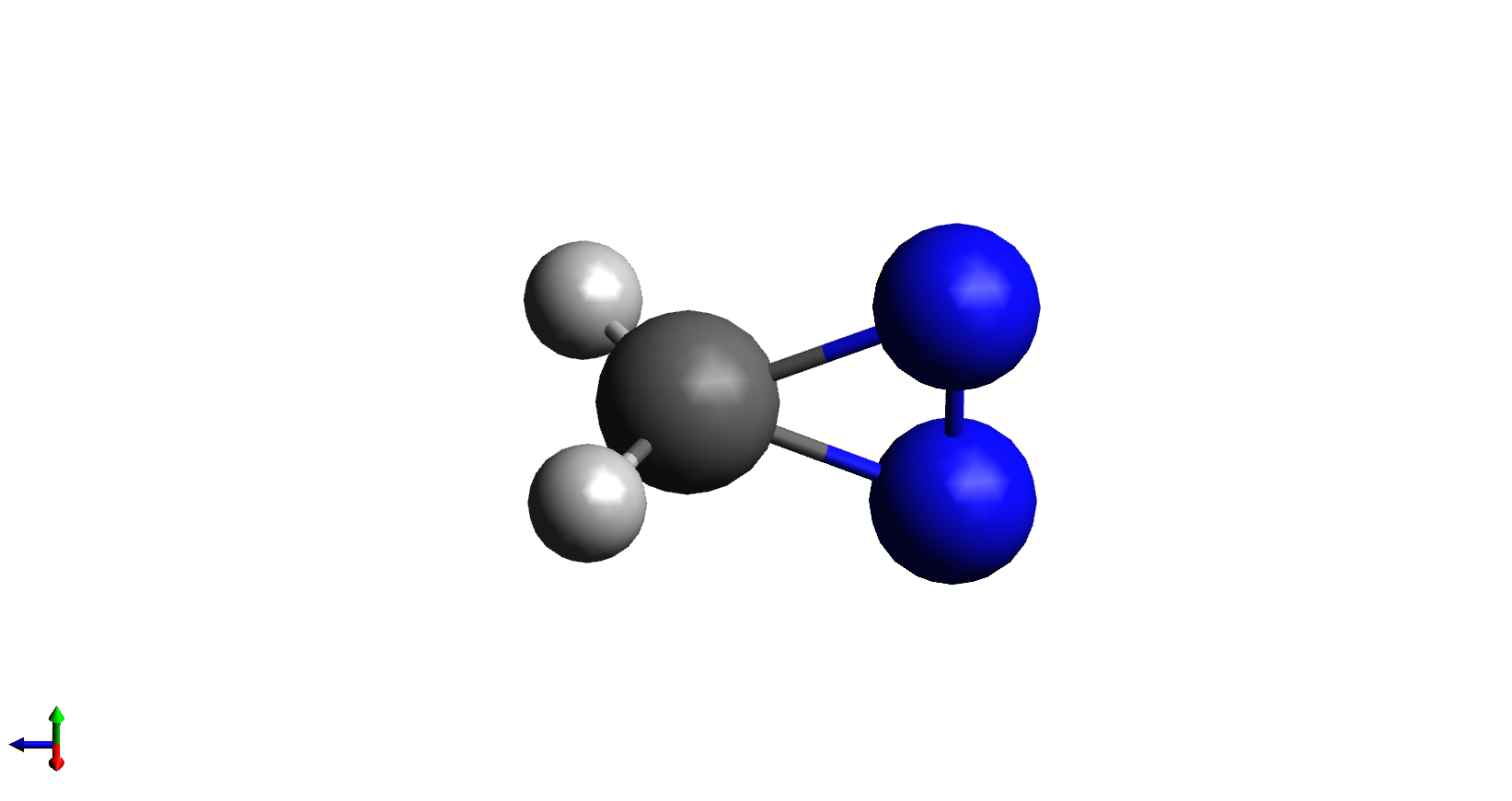}
\end{center}

\subsubsection*{Optimized Atomic Coordinates in Å}
\begin{tabular}{lrrr}
Element & x & y & z \\

C & 0.00000007890147 & -0.00000013947960 & 0.88998932718176 \\

N & -0.00000001940403 & 0.60228246380871 & -0.48374761824502 \\

N & -0.00000001938792 & -0.60228249992511 & -0.48374785483074 \\

H & 0.92881356190568 & 0.00000008779278 & 1.41750408927633 \\

H & -0.92881360201520 & 0.00000008780322 & 1.41750405661767 \\

\end{tabular}

\subsubsection*{Vibrational Frequencies (cm$^{-1}$)}
\begin{tabular}{ cccccccc }

    0.00 & 0.00 & 0.00 & 0.00 & 0.00 & 0.00 & 838.79 & 996.41 \\

    1016.27 & 1076.79 & 1174.81 & 1599.02 & 1821.08 & 3310.39 & 3449.74 \\

\end{tabular}

\subsection{Structure: TS0}
\textbf{Level of Theory:} CASSCF (12,10)


\begin{center}
    \includegraphics[trim={5cm 0 0 0},clip,width=0.3\textwidth]{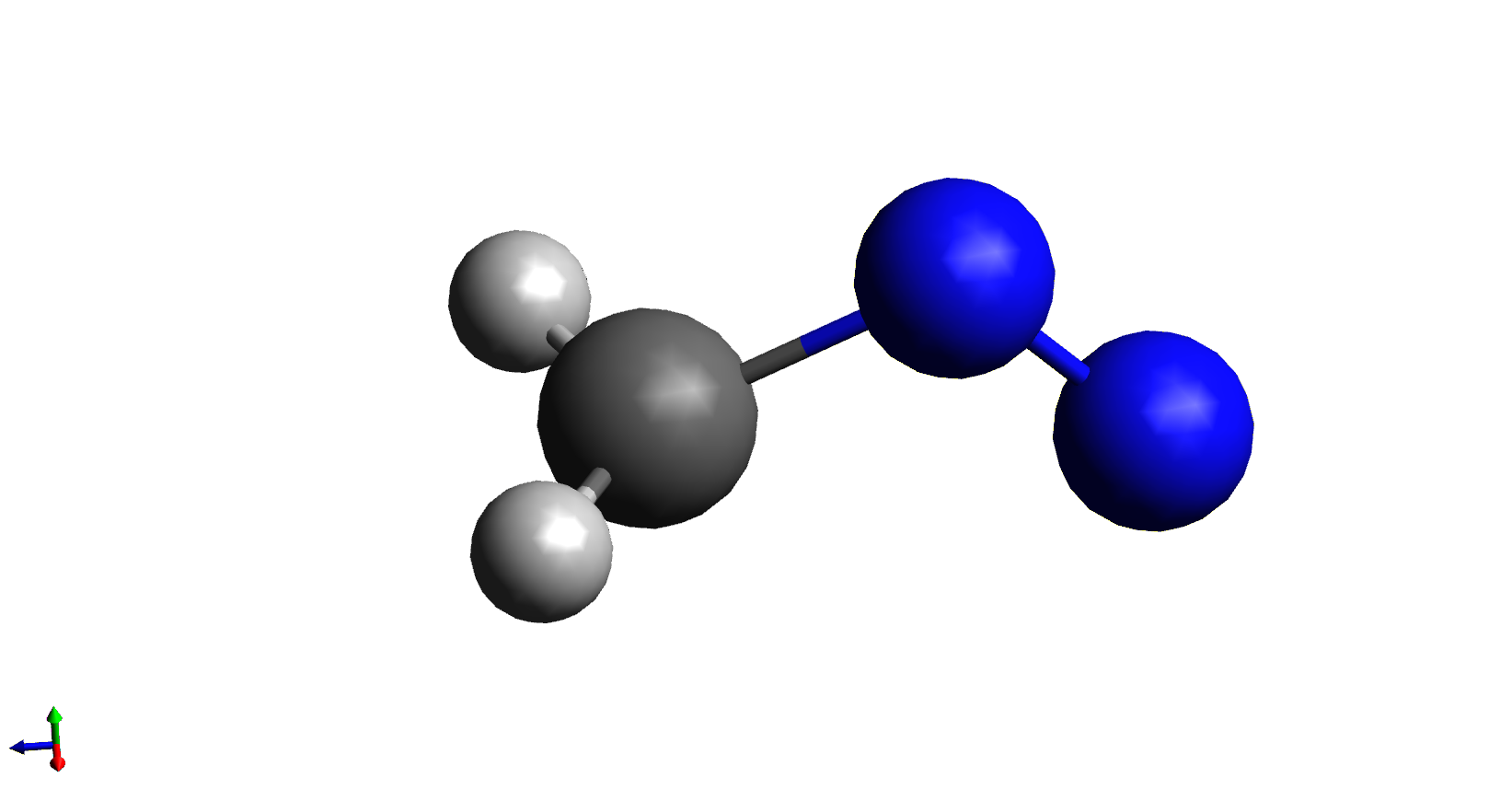}
\end{center}

\subsubsection*{Optimized Atomic Coordinates in Å}
\begin{tabular}{lrrr}
Element & x & y & z \\

C & -0.00012019083262 & -0.00162451405831 & 0.87983704101779 \\

N & -0.00040484150445 & 0.61256727837446 & -0.47220402050980 \\

N & 0.00046827874955 & -0.25738266225190 & -1.28165669580996 \\

H & 0.94332928020216 & -0.00118582859049 & 1.38265090499403 \\

H & -0.94327364354781 & -0.00078295154415 & 1.38317728562233 \\

\end{tabular}

\subsubsection*{Vibrational Frequencies (cm$^{-1}$)}
\begin{tabular}{ cccccccc }

    0.00 & 0.00 & 0.00 & 0.00 & 0.00 & 0.00 & -903.29 & 156.14 \\

    760.83 & 985.32 & 1058.90 & 1529.45 & 1789.26 & 3322.53 & 3491.64 \\

\end{tabular}

\subsection{Structure: TS1}
\textbf{Level of Theory:} CASSCF (12,10)


\begin{center}
    \includegraphics[trim={5cm 0 0 0},clip,width=0.4\textwidth]{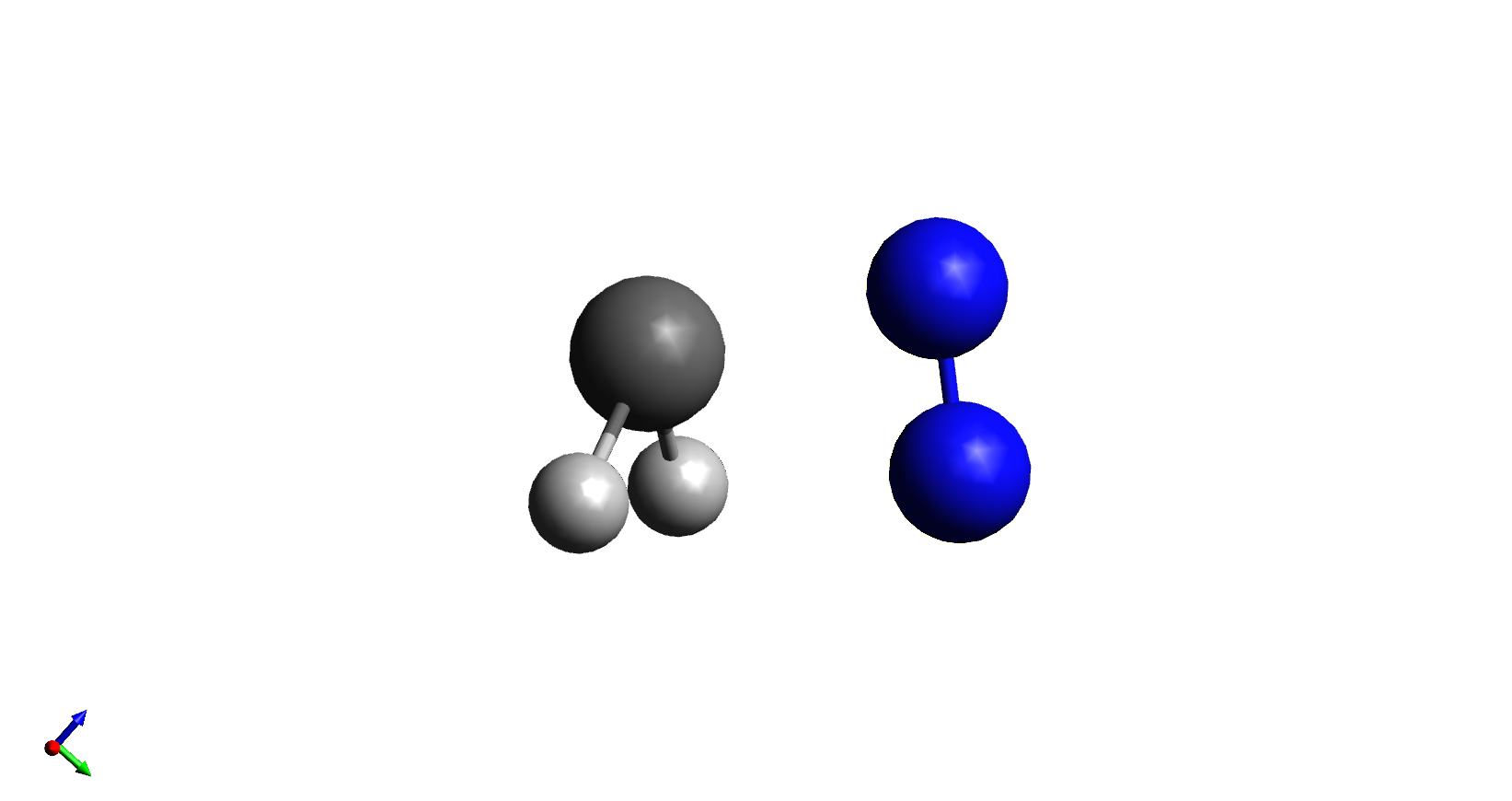}
\end{center}

\subsubsection*{Optimized Atomic Coordinates in Å}
\begin{tabular}{lrrr}
Element & x & y & z \\

C & -0.00240097781410 & -0.01022555957572 & -0.00356087161745 \\

N & 0.00093234403710 & 1.88641905556670 & 0.72865449672662 \\

N & 0.00102813030034 & 1.04632740838791 & 1.47883678019754 \\

H & 0.89808370260560 & 0.25090169210550 & -0.54202987206299 \\

H & -0.89764319912894 & 0.25955582016160 & -0.54648510485772 \\

\end{tabular}

\subsubsection*{Vibrational Frequencies (cm$^{-1}$)}
\begin{tabular}{ cccccccc }

    0.00 & 0.00 & 0.00 & 0.00 & 0.00 & 0.00 & -580.07 & 450.77 \\

    542.05 & 1099.10 & 1133.66 & 1469.07 & 2122.14 & 3196.49 & 3331.60 \\

\end{tabular}

\subsection{Structure: A2DM}
\textbf{Level of Theory:} CASSCF (12,10)


\begin{center}
    \includegraphics[trim={5cm 0 0 0},clip,width=0.3\textwidth]{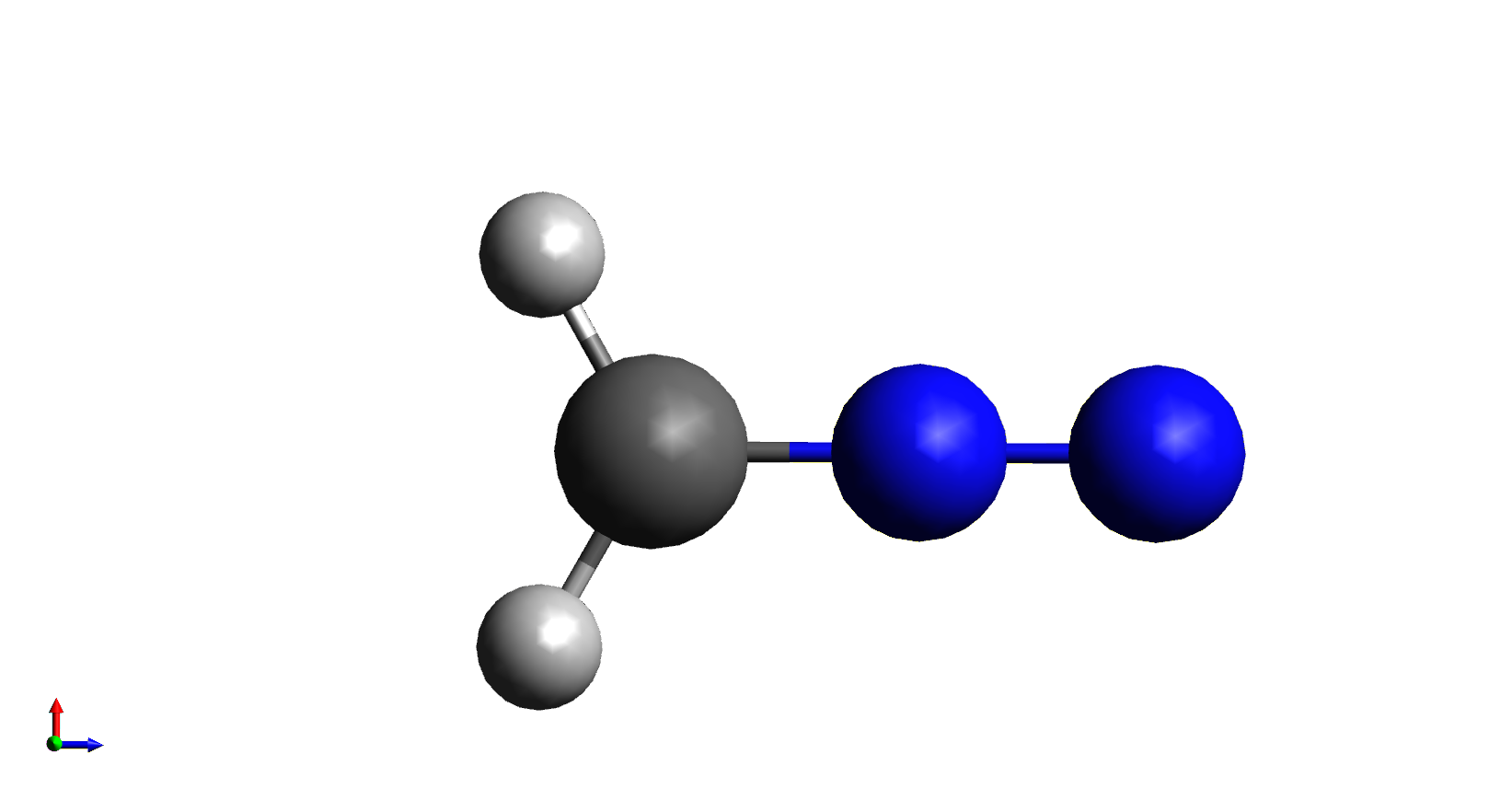}
\end{center}

\subsubsection*{Optimized Atomic Coordinates in Å}
\begin{tabular}{lrrr}
Element & x & y & z \\

C & -0.00000000002039 & -0.00000000000017 & -1.21024361045821 \\

N & -0.00000000004924 & 0.00000000000004 & 0.09339179739379 \\

N & 0.00000000006048 & 0.00000000000000 & 1.24965184836771 \\

H & 0.95471709497004 & 0.00000000000006 & -1.75048201761099 \\

H & -0.95471709496090 & 0.00000000000006 & -1.75048201769230 \\

\end{tabular}

\subsubsection*{Vibrational Frequencies (cm$^{-1}$)}
\begin{tabular}{ cccccccc }

    0.00 & 0.00 & 0.00 & 0.00 & 0.00 & 0.00 & -265.26 & 443.23 \\

    539.12 & 1154.84 & 1172.39 & 1487.36 & 2151.23 & 2992.02 & 3088.12 \\

\end{tabular}

\subsection{Structure: B1DZ}
\textbf{Level of Theory:} CASSCF (12,10)


\begin{center}
    \includegraphics[trim={5cm 0 0 0},clip,width=0.3\textwidth]{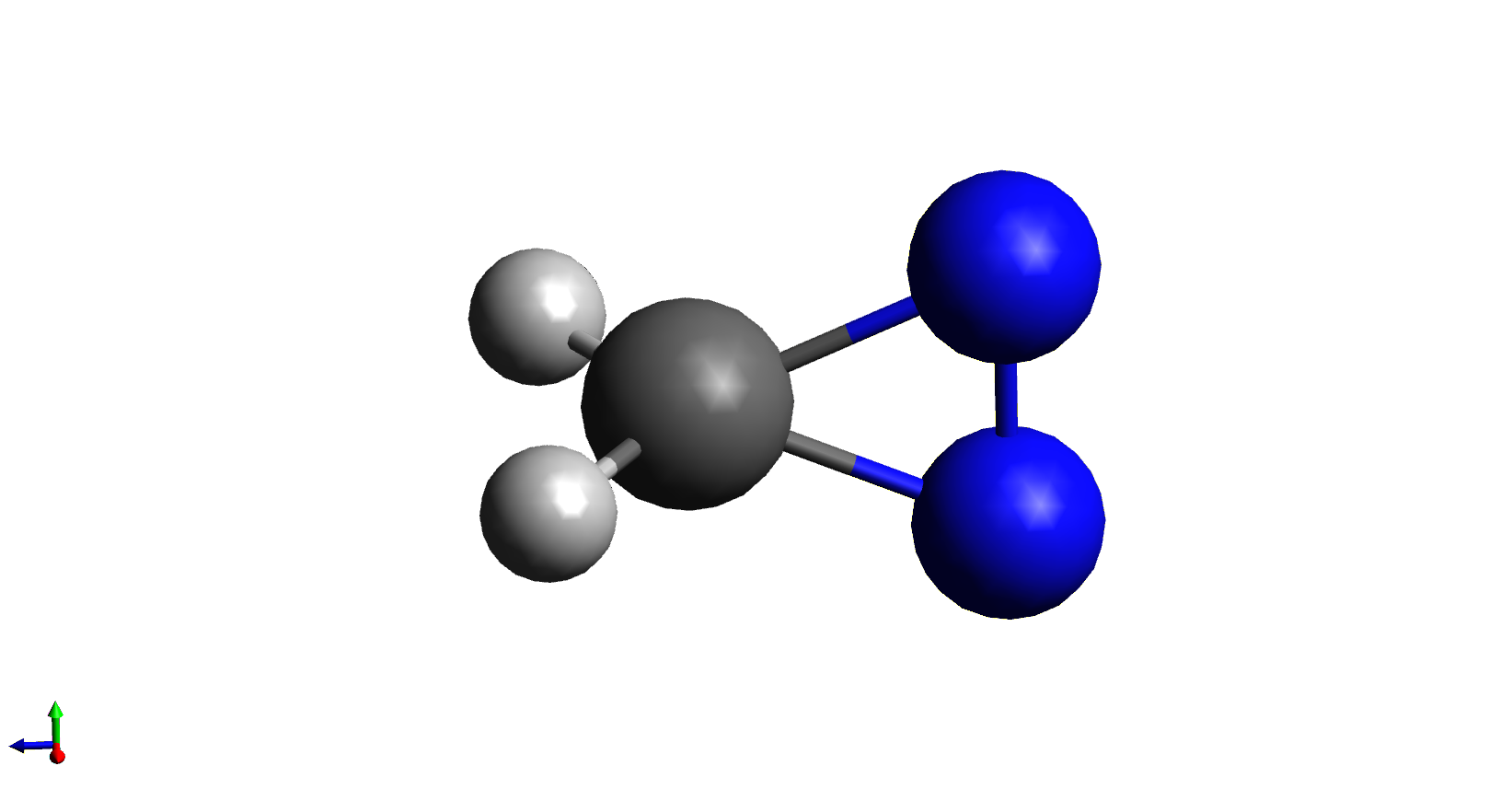}
\end{center}

\subsubsection*{Optimized Atomic Coordinates in Å}
\begin{tabular}{lrrr}
Element & x & y & z \\

C & -0.00000007082969 & -0.00000255456676 & 0.91144501083161 \\

N & 0.00000000176061 & 0.61016990693536 & -0.50694048808469 \\

N & 0.00000000283861 & -0.61016926762071 & -0.50694069941617 \\

H & 0.93266169476923 & 0.00000095727106 & 1.42996909984116 \\

H & -0.93266162853876 & 0.00000095798104 & 1.42996907682809 \\

\end{tabular}

\subsubsection*{Vibrational Frequencies (cm$^{-1}$)}
\begin{tabular}{ cccccccc }

    0.00 & 0.00 & 0.00 & 0.00 & 0.00 & 0.00 & 614.56 & 792.51 \\

    893.02 & 1103.98 & 1149.37 & 1595.68 & 1789.62 & 3332.39 & 3487.02 \\

\end{tabular}

\subsection{Structure: CI1}
\textbf{Level of Theory:} CASSCF (12,10)


\begin{center}
    \includegraphics[trim={5cm 0 0 0},clip,width=0.4\textwidth]{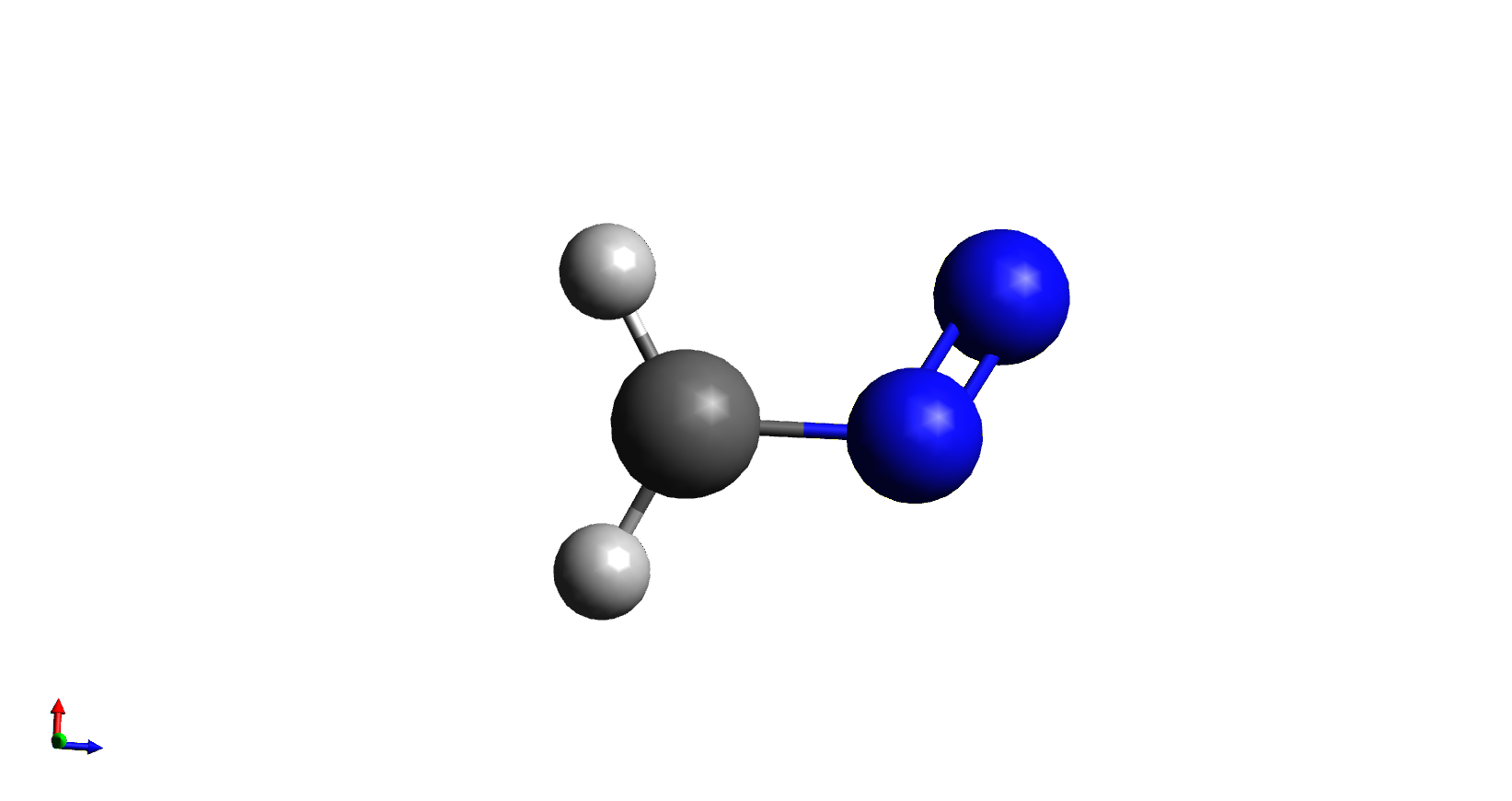}
\end{center}

\subsubsection*{Optimized Atomic Coordinates in Å}
\begin{tabular}{lrrr}
Element & x & y & z \\

C & 0.01069461295727 & 0.09018251549151 & -0.01742593668652 \\

N & 0.02583150339802 & 0.16461163435570 & 1.42317575951757 \\

N & 1.00208719631919 & -0.24587278584475 & 1.95835346844432 \\

H & 0.92508162703853 & 0.21946962808588 & -0.57513990411368 \\

H & -0.91581293971304 & -0.22839199208834 & -0.46696738716175 \\

\end{tabular}

\subsubsection*{Vibrational Frequencies (cm$^{-1}$)}
\begin{tabular}{ cccccccc }

    0.00 & 0.00 & 0.00 & 0.00 & 0.00 & 0.00 & -604.19 & -80.10 \\

    846.76 & 1067.80 & 1404.92 & 1622.68 & 3170.72 & 3332.49 & 4877.17 \\

\end{tabular}

\newpage
\subsection{Structure: PhDM}
\textbf{Level of Theory:} RI-MP2


\begin{center}
    \includegraphics[trim={5cm 0 0 0},clip,width=0.4\textwidth]{ Figures/Ph\_a1dm.png }
\end{center}

\subsubsection*{Optimized Atomic Coordinates in Å}
\begin{tabular}{lrrr}
Element & x & y & z \\

C & -1.50447956037168 & -1.43504804055473 & -0.00306211947524 \\

C & -0.19512945836494 & -0.96689727779058 & -0.00615329284035 \\

C & 0.06958587874276 & 0.41136160270097 & -0.00057011018126 \\

C & -1.01573064537067 & 1.30173280413338 & 0.00831960674793 \\

C & -2.32164392930904 & 0.82705428386479 & 0.01131059442691 \\

C & -2.57580719409660 & -0.54386619685359 & 0.00571390481812 \\

H & -1.68625565066297 & -2.50112280608754 & -0.00724689215459 \\

H & 0.62362747973397 & -1.67622563064643 & -0.01314703450226 \\

H & -0.82821420107010 & 2.36786073464823 & 0.01264675732050 \\

H & -3.14289199157368 & 1.53073639676741 & 0.01832677857485 \\

H & -3.59214225705642 & -0.91124963424002 & 0.00781488707561 \\

C & 1.41443536029562 & 0.94428589233630 & -0.00347417712167 \\

H & 1.67088490215722 & 1.98950672897545 & -0.00012367061352 \\

N & 2.43801261992123 & 0.11679143777032 & -0.01227249843064 \\

N & 3.29693234702531 & -0.63088059502397 & -0.02026463364440 \\

\end{tabular}

\subsubsection*{Vibrational Frequencies (cm$^{-1}$)}
\begin{tabular}{ cccccccc }

    0.00 & 0.00 & 0.00 & 0.00 & 0.00 & 0.00 & 59.66 & 118.92 \\

    209.87 & 332.85 & 354.61 & 401.42 & 460.37 & 499.15 & 522.63 & 620.78 \\

    659.08 & 679.07 & 741.55 & 835.12 & 847.18 & 882.44 & 947.71 & 957.47 \\

    1009.84 & 1049.67 & 1101.18 & 1150.27 & 1177.44 & 1202.16 & 1223.56 & 1342.36 \\

    1404.99 & 1457.42 & 1489.97 & 1527.61 & 1618.82 & 1646.33 & 2369.75 & 3193.29 \\

    3200.03 & 3212.49 & 3222.04 & 3236.22 & 3277.38 \\

\end{tabular}
\newpage

\subsection{Structure: PhDZ}
\textbf{Level of Theory:} RI-MP2


\begin{center}
    \includegraphics[trim={5cm 0 0 0},clip,width=0.4\textwidth]{ Figures/Ph\_a1dz.png }
\end{center}

\subsubsection*{Optimized Atomic Coordinates in Å}
\begin{tabular}{lrrr}
Element & x & y & z \\

C & 2.14388500180402 & 0.97772362295494 & -0.00008368455532 \\

C & 0.79319503607885 & 1.30790119548344 & -0.00006761998838 \\

C & -0.18125164480261 & 0.30438377906803 & 0.00001803700535 \\

C & 0.21676041526792 & -1.03636762109846 & 0.00008940904298 \\

C & 1.57008657746334 & -1.36310270539698 & 0.00006933064800 \\

C & 2.53676641462681 & -0.36013854849450 & -0.00001801774766 \\

H & 2.88859305244261 & 1.76155706825682 & -0.00013848084514 \\

H & 0.48839967618577 & 2.34673517410645 & -0.00010641646436 \\

H & -0.52691443429259 & -1.82142657143851 & 0.00017145290690 \\

H & 1.86864050868580 & -2.40229437077265 & 0.00016365787043 \\

H & 3.58660170792747 & -0.61753503747503 & -0.00003091492996 \\

C & -1.59788155334928 & 0.67729490008217 & 0.00008978025319 \\

H & -1.88076905663159 & 1.71857186916536 & 0.00057273369582 \\

N & -2.57638480389722 & -0.25389256384428 & -0.62262953469619 \\

N & -2.57649309750931 & -0.25453209059681 & 0.62163456780435 \\

\end{tabular}

\subsubsection*{Vibrational Frequencies (cm$^{-1}$)}
\begin{tabular}{ cccccccc }

    0.00 & 0.00 & 0.00 & 0.00 & 0.00 & 0.00 & 68.97 & 163.65 \\

    192.88 & 362.85 & 389.05 & 406.06 & 550.27 & 601.53 & 620.00 & 696.87 \\

    757.77 & 819.46 & 823.97 & 848.45 & 909.63 & 957.50 & 971.11 & 1008.44 \\

    1013.29 & 1016.18 & 1055.80 & 1109.06 & 1181.96 & 1204.40 & 1262.65 & 1337.26 \\

    1370.77 & 1466.46 & 1483.75 & 1524.54 & 1559.44 & 1630.41 & 1652.71 & 3203.94 \\

    3211.33 & 3221.24 & 3228.75 & 3230.25 & 3240.20 \\

\end{tabular}
\newpage

\subsection{Structure: PhTS0}
\textbf{Level of Theory:} RI-MP2


\begin{center}
    \includegraphics[trim={5cm 0 0 0},clip,width=0.4\textwidth]{ Figures/Ph\_TS3.png }
\end{center}

\subsubsection*{Optimized Atomic Coordinates in Å}
\begin{tabular}{lrrr}
Element & x & y & z \\

C & 2.14196393220928 & 0.95321366866322 & 0.27546854946023 \\

C & 0.83735549944883 & 1.28839359970607 & -0.06864276674659 \\

C & -0.08782479848092 & 0.29545709013145 & -0.44284651078901 \\

C & 0.33363233736522 & -1.04724792577399 & -0.44176013718036 \\

C & 1.62399530870790 & -1.38814871574640 & -0.05923443056613 \\

C & 2.53136581615601 & -0.38630393067079 & 0.29151131568384 \\

H & 2.84997909032809 & 1.72457513930017 & 0.54543856627439 \\

H & 0.51823997932221 & 2.32290514024461 & -0.06076307449279 \\

H & -0.37477215228702 & -1.80170391392972 & -0.75727682945330 \\

H & 1.93521796277370 & -2.42356349474617 & -0.05282655683503 \\

H & 3.54231084236111 & -0.64929333997135 & 0.57216379779925 \\

C & -1.42361875129461 & 0.60133656918960 & -0.95836876589696 \\

H & -1.59482011241058 & 1.68267676207823 & -0.84422687661797 \\

N & -2.56625162638677 & 0.23008441144464 & 0.64016930011513 \\

N & -3.28251632781246 & -0.43479025991959 & 1.18502651924531 \\

\end{tabular}

\subsubsection*{Vibrational Frequencies (cm$^{-1}$)}
\begin{tabular}{ cccccccc }

    0.00 & 0.00 & 0.00 & 0.00 & 0.00 & 0.00 & -288.65 & 32.12 \\

    42.49 & 143.88 & 219.21 & 304.07 & 353.25 & 409.64 & 487.56 & 549.94 \\

    618.66 & 693.58 & 767.59 & 833.47 & 858.41 & 933.17 & 970.19 & 973.30 \\

    982.52 & 1012.15 & 1047.46 & 1080.11 & 1130.78 & 1181.18 & 1190.54 & 1257.75 \\

    1334.02 & 1461.49 & 1479.19 & 1507.89 & 1613.93 & 1631.94 & 2117.94 & 3030.26 \\

    3201.15 & 3209.00 & 3217.26 & 3227.44 & 3235.86 \\

\end{tabular}
\newpage

\subsection{Structure: PhTS1}
\textbf{Level of Theory:} RI-MP2


\begin{center}
    \includegraphics[trim={5cm 0 0 0},clip,width=0.4\textwidth]{ Figures/Ph\_TS1.png }
\end{center}

\subsubsection*{Optimized Atomic Coordinates in Å}
\begin{tabular}{lrrr}
Element & x & y & z \\

C & 2.08963088770727 & 0.98675347937290 & 0.27978701934553 \\

C & 0.75556031632492 & 1.26237731616116 & 0.00203382334880 \\

C & -0.13242423619665 & 0.22680838508913 & -0.34369088415767 \\

C & 0.34754019127193 & -1.09586005001731 & -0.37192876435467 \\

C & 1.66613329005071 & -1.37492071144868 & -0.04274445017058 \\

C & 2.54189757286609 & -0.33250837365374 & 0.27039908938015 \\

H & 2.77086364351591 & 1.78952232525583 & 0.52555790284332 \\

H & 0.38496617812888 & 2.27845804151238 & 0.04203347938989 \\

H & -0.33472247818566 & -1.88513997420765 & -0.65854632422728 \\

H & 2.02448185267746 & -2.39484305396846 & -0.05269770230244 \\

H & 3.57499849781405 & -0.54918337565552 & 0.50555882907319 \\

C & -1.46869825748411 & 0.50259450766658 & -0.84059262693911 \\

H & -1.77577430657202 & 1.54393677430667 & -0.68830529776434 \\

N & -2.74902030720521 & -0.41967156666126 & 0.24114813392706 \\

N & -2.59210214471359 & 0.39662617624796 & 1.02929507260815 \\

\end{tabular}

\subsubsection*{Vibrational Frequencies (cm$^{-1}$)}
\begin{tabular}{ cccccccc }

    0.00 & 0.00 & 0.00 & 0.00 & 0.00 & 0.00 & -448.01 & 63.17 \\

    97.00 & 165.35 & 249.60 & 342.40 & 399.11 & 444.00 & 487.76 & 553.91 \\

    618.30 & 690.49 & 763.20 & 839.39 & 852.92 & 922.27 & 941.89 & 970.26 \\

    978.03 & 1011.55 & 1046.91 & 1097.46 & 1164.57 & 1182.62 & 1195.69 & 1271.24 \\

    1336.91 & 1460.04 & 1482.15 & 1509.47 & 1611.66 & 1632.47 & 1903.05 & 3088.48 \\

    3204.51 & 3212.29 & 3220.67 & 3228.82 & 3237.81 \\

\end{tabular}
\newpage

\section{2-Qubit gate counts and SQD sub-space dimensions}

\begin{table}[ht!]

\centering
\caption{\textbf{Average subspace dimensions for diagonalization in Sample-based Quantum Diagonalization (SQD) method and 2-qubit CZ gate count post transpilation to quantum hardware for the parent and phenyl-substituted system.}}
\begin{tabular}{lcc}

\hline
\multicolumn{3}{c}{\textbf{DZ-DM (12,10)}} \\
\hline
Structure & Avg. Subspace Dim & $\#$2-Qubit Gates \\
\hline
A1DM & 28654 & 292\\
A1DZ & 32454 & 296\\
A2DM & 29864 & 300\\
B1DZ & 31763 & 292\\
CI1 & 32827 & 314\\
TS0 & 31639 & 314\\
TS1 & 34162 & 314\\
\hline
\multicolumn{3}{c}{\textbf{PhDZ-PhDM (30,30)}} \\
\hline
Structure & Avg. Subspace Dim & $\#$2-Qubit Gates \\
\hline
PhDM & $2.25*10^{8}$ & 2944\\
PhDZ & $2.25*10^{8}$ & 2914 \\
PhTS1 & $2.25*10^{8}$ & 2983\\
PhTS0 & $2.25*10^{8}$ & 2990\\
\hline
\end{tabular}

\end{table}
\newpage
\section{Energy}

\begin{table}[ht!]
\centering
\caption{\textbf{Energies (Hartrees) for DZ-DM structures computed using various quantum chemistry methods.}}
\begin{threeparttable}
\resizebox{\textwidth}{!}{
\begin{tabular}{lrrrrrr}
\toprule
\multicolumn{7}{c}{\textbf{DZ-DM}} \\
\midrule
Structure & DFT & CCSD & CASSCF \small(12,10)\tnote{a} & CASSCF \small(12,10) & CASCI \small(12,10) & SQD \small(12,10) \\
\midrule
A1DM & -148.7176519 & -148.4679104 & -148.07326   & -148.0326574 & -147.9639947 & -147.9639832 \\
A1DZ & -148.691536  & -148.4547860  & -148.05365   & -148.0193985 & -147.9270850  & -147.9270840 \\
TS0  & -148.591058  & -148.3521807 & -147.98248   & -147.9355890  & -147.8702703 & -147.8702548 \\
TS1  & -148.6459075 & -148.4050091 & -148.01573   & -147.9839341 & -147.9141154 & -147.9141072 \\
\midrule
Structure & TD-DFT & EOM-CCSD & CASSCF \small(12,10)\tnote{a} & SA-CASSCF \small(12,10) & SA-CASCI \small(12,10) & Ext-SQD \small(12,10)\\
\midrule
A2DM & -148.6102461 & -148.3537078 & -147.95614   & -147.9193937 & -147.8395428 & -147.8395428 \\
B1DZ & -148.5494472 & -148.3004624 & -147.88929   & -147.8503496 & -147.7551589 & -147.7551589 \\
CI1  & -148.6234315 & -148.3616668\tnote{b} & -147.94963 & -147.9518018 & -147.8699035 & -147.8659217 \\
\bottomrule
\end{tabular}
}
\begin{tablenotes}
\tiny
\item [a] CASSCF(12,10) data from Arenas et al.~\cite{arenas2002carbene}. Note that the active space differs from the one used in the present study.
\item [b] CCSD value for CI1 obtained using the CCSD2 implementation in ORCA.
\end{tablenotes}
\end{threeparttable}
\label{tab:energies_DM_DZ}
\end{table}

\begin{table}[h!]
\centering
\caption{\textbf{Energies (Hartrees) for PhDZ-PhDM structures computed using various quantum chemistry methods.}}
\begin{threeparttable}
\resizebox{\textwidth}{!}{
\begin{tabular}{lrrrrrr}
\toprule
\multicolumn{7}{c}{\textbf{PhDZ-PhDM}} \\
\midrule
Structure & DFT & CCSD & RI-CC2/TZVP\tnote{a} & CCSD \small(30,30) & SCI \small(30,30) & SQD \small(30,30) \\
\midrule
A1DM & -379.7085334 & -379.1762876 & -378.8349480  & -377.8578443  & -377.8424989 & -377.7663256 \\
A1DZ & -379.6819668 & -379.1626667 & -377.4879870  & -377.8515668 & -377.8303039 & -377.7532508 \\
PhTS0  & -379.6490675 & -379.1243250  & -378.7808680  & -377.8341738 & -377.8156839 & -377.7360293 \\
PhTS1  & -379.6373094 & -379.1080457 & -378.7680863 & -377.8100869 & -377.7906091 & -377.7136366 \\
\bottomrule
\end{tabular}
}
\begin{tablenotes}
\tiny
\item [a] RI-CC2/TZVP data from Yunlong et al.~\cite{phenyldz_abinitio}.
\end{tablenotes}
\end{threeparttable}
\label{tab:energies_PhDM_PhDZ}
\end{table}

